\documentclass[%
 showpacs,
 amsmath,amssymb,
]{revtex4-1}

\usepackage{graphicx}
\usepackage{dcolumn}
\usepackage{bm}

\usepackage{CJK}

\usepackage{color}
\usepackage{amssymb}
\usepackage{amsfonts}

\usepackage[colorlinks,urlcolor=blue,linkcolor=blue,citecolor=blue,anchorcolor=blue]{hyperref}

\begin{document}

\preprint{APS/123-QED}

\title{Dual accelerating Airy-Talbot recurrence effect}

\author{Yiqi Zhang$^1$}
\email{zhangyiqi@mail.xjtu.edu.cn}
\author{Hua Zhong$^1$}
\author{Milivoj R. Beli\'c$^{2}$}
\author{Xing Liu$^1$}
\author{Weiping Zhong$^3$}
\author{Yanpeng Zhang$^{1}$}
\email{ypzhang@mail.xjtu.edu.cn}
\author{Min Xiao$^{4,5}$}
\affiliation{%
 $^1$Key Laboratory for Physical Electronics and Devices of the Ministry of Education \& Shaanxi Key Lab of Information Photonic Technique,
Xi'an Jiaotong University, Xi'an 710049, China \\
$^2$Science Program, Texas A\&M University at Qatar, P.O. Box 23874 Doha, Qatar \\
$^3$Department of Electronic and Information Engineering, Shunde Polytechnic, Shunde 528300, China \\
$^4$Department of Physics, University of Arkansas, Fayetteville, Arkansas, 72701, USA \\
$^5$National Laboratory of Solid State Microstructures and School of Physics, Nanjing University, Nanjing 210093, China
}%

\date{\today}

\begin{abstract}
  \noindent
 We demonstrate the dual accelerating Airy-Talbot recurrence effect,
  i.e., the self-imaging of accelerating optical beams, by propagating a superposition of Airy beams with successively changing transverse displacements.
  The dual Airy-Talbot effect is a spontaneous recurring imaging of the input and of the input with alternating component signs.
  It results from the constructive interference of Airy wave functions, which is also responsible for other kinds of Airy beams, for example, Airy breathers.
  An input composed of finite-energy Airy beams also displays the dual Airy-Talbot effect, but it demands a large transverse displacement
  and diminishes fast along the propagation direction.
\end{abstract}

\maketitle

%

\noindent
Talbot effect is a self-imaging phenomenon,
first observed by H. F. Talbot \cite{talbot.pm.9.401.1836}
and explained by Lord Rayleigh \cite{rayleigh.pm.11.196.1881}.
Owing to its potential applications in image preprocessing and synthesis, photolithography,
optical testing, optical metrology, spectrometry, and optical computing,
Talbot effect has been reported in, but not confined to,
atomic optics \cite{wen.apl.98.081108.2011,zhang.ieee.4.2057.2012},
quantum optics \cite{song.prl.107.033902.2011,jin.apl.101.211115.2012},
waveguide arrays \cite{iwanow.prl.95.053902.2005},
photonic lattices \cite{ramezani.prl.109.033902.2012},
Bose-Einstein condensates \cite{deng.prl.83.5407.1999,ryu.prl.96.160403.2006},
and X-ray imaging \cite{pfeiffer.nm.7.134.2008}.
Talbot effect can also be obtained using
harmonic waves \cite{zhang.prl.104.183901.2010}, spherical waves \cite{azana.prl.112.213902.2014} and rogue waves
\cite{zhang.pre.89.032902.2014,zhang.pre.91.032916.2015}, but a general requirement for its appearance is the transverse periodicity of the input wave.
Thus, an infinitely-extending transversely-periodic input will generate an infinitely-extending longitudinally-periodic self-imaging output -- a Talbot carpet.
For finite-window periodic or nearly-periodic inputs, partial reconstructions are possible in the near-field, due to wave interference.
However, a longitudinal self-imaging may not be caused by transversely-periodic beam structures only \cite{turunen.jossa.8.282.1991,lumer.prl.115.013901.2015} -- but this then is not the Talbot effect in the usual sense.
For a thorough reading on Talbot effect, readers are directed to Ref. \cite{wen.aop.5.83.2013} and references therein.

Very recently, Airy-Talbot effect, formed from a superposition of Airy functions, was introduced \cite{lumer.prl.115.013901.2015}.
This effect is quite different from previous research on Talbot, because the input is an asymmetric self-accelerating beam, and consequently, the self-images also accelerate during propagation.
In the last decade, Airy and related accelerating beams have attracted widespread attention \cite{berry.ajp.47.264.1979,siviloglou.ol.32.979.2007,siviloglou.prl.99.213901.2007},
and have been reported in nonlinear media \cite{kaminer.prl.106.213903.2011,lotti.pra.84.021807.2011,dolev.prl.108.113903.2012,zhang.ol.38.4585.2013,zhang.oe.22.7160.2014,shen.sr.5.9814.2015},
Bose-Einstein condensates \cite{efremidis.pra.87.043637.2013},
on the surface of a metal \cite{salandrino.ol.35.2082.2010,zhang.ol.36.3191.2011,minovich.prl.107.116802.2011,li.prl.107.126804.2011},
in the harmonic potential \cite{zhang.oe.23.10467.2015,zhang.ol.40.3786.2015}, and elsewhere.

In this Letter, 
we demonstrate the dual Airy-Talbot effect, theoretically and numerically.
The dual Airy-Talbot effect does not contain the images of the input only;
the superpositions of Airy beams with successively increasing transverse displacements form two types of sharp images --
of the input and of the input with alternating signs of the components.
Hence the name, the dual Airy-Talbot effect. The effect is carried over to the finite-energy Airy beams, but with diminishing range and clarity.
We establish that effectively the Airy-Talbot effect and the Airy breathers have a common cause -- the constructive interference of linear accelerating beams.

In free space, the paraxial wave equation can be written as the dimensionless Schr\"odinger equation
\begin{equation}\label{eq1}
  i\frac{\partial \psi}{\partial z} + \frac{1}{2}\frac{\partial^2\psi}{\partial x^2}=0,
\end{equation}
where $x$ and $z$ are the normalized transverse coordinate and the propagation distance, scaled by some characteristic
transverse width and the corresponding Rayleigh range \cite{zhang.ol.38.4585.2013,zhang.oe.22.7160.2014}.
It is astonishing that such a simple parabolic partial differential equation can still offer interesting dynamics of its specific particular solutions.
According to the Fresnel diffraction integral \cite{goodman.book.2005}, for any input $\psi(x)$
the corresponding propagating solution can be written as
\begin{equation}\label{eq2}
  \psi(x,z)=\sqrt{\frac{-i}{2\pi z}} \int_{-\infty}^{+\infty} \psi(\xi) \exp\left[\frac{i}{2z}(x-\xi)^2\right] d\xi.
\end{equation}
We assume that the input $\psi$ is chosen as a sum of equidistant Airy functions
\begin{equation}\label{eq3}
  \psi(x)=\sum_{n\in\mathbb{Z}}c_n {\rm Ai}(x-n\delta),
\end{equation}
where $c_n$ is an arbitrary amplitude coefficient, and $\delta$ is an arbitrary transverse displacement.
Plugging Eq. (\ref{eq3}) into Eq. (\ref{eq2}), one obtains
\begin{align}\label{eq4}
  \psi(x,z) = \sqrt{\frac{-i}{2\pi z}} & \exp\left(\frac{i}{2z}x^2\right) \sum_{n\in\mathbb{Z}}c_n \int_{-\infty}^{+\infty} {\rm Ai}(\xi-n\delta) 
   \exp\left(\frac{i}{2z}\xi^2\right) \exp\left(-\frac{i}{z}x\xi\right) d\xi.
\end{align}
The integral in Eq. (\ref{eq4}) can be rewritten as a convolution of the Fourier transforms of ${\rm Ai}(x-n\delta)$ and $\exp({ix^2}/{2z})$ with $x/z$ being the spatial frequency.
As a result, Eq. (\ref{eq4}) takes the form
\begin{align}\label{eq5}
  \psi(x,z) = & \sqrt{\frac{-i}{2\pi z}} \exp\left(\frac{i}{2z}x^2\right) \sum_{n\in\mathbb{Z}} c_n\sqrt{i2\pi z}
   \int_{-\infty}^{+\infty} \exp\left(\frac{i}{3}\frac{\xi^3}{z^3}-i\frac{\xi}{z}n\delta\right) \exp\left[-\frac{i}{2z}(x-\xi)^2\right] d\xi .
\end{align}
After some algebra, Eq. (\ref{eq5}) can be rewritten as
\begin{align}\label{eq6}
  \psi = \sum_{n\in\mathbb{Z}}\frac{c_n}{i}  \int_{-i\infty}^{+i\infty} \exp\left[-\frac{\xi^3}{3}+(x-n\delta)\xi\right] \exp\left(i\frac{z}{2}\xi^2\right) d\xi.
\end{align}
If we let $\xi=\eta+iz/2$, Eq. (\ref{eq6}) is recast into
\begin{align}\label{eq7}
  \psi(x,z) =  \sum_{n\in\mathbb{Z}}\frac{c_n}{i}  \exp\left[i\frac{z}{2}\left(x-n\delta-\frac{z^2}{6}\right)\right] 
  \left\{\int_{-i\infty}^{+i\infty} \exp\left[-\frac{\eta^3}{3}+\left(x-n\delta-\frac{z^2}{4}\right)\eta \right] d\eta \right\}.
\end{align}
By using the definition of Airy function \cite{vallee.book.2010}, one can write
\begin{equation}\label{eq8}
  \psi=\sum_{n\in\mathbb{Z}}c_n {\rm Ai}\left(x-n\delta-\frac{z^2}{4}\right) \exp\left[i\frac{z}{2}\left(x-n\delta-\frac{z^2}{6}\right)\right],
\end{equation}
which is exactly Eq. (4) -- the solution obtained in Ref. \cite{lumer.prl.115.013901.2015}.
Naturally, the corresponding intensity is
\begin{equation}\label{eq9}
  I(x,z)=\left|\sum_{n\in\mathbb{Z}}c_n {\rm Ai}\left(x-n\delta-\frac{z^2}{4}\right) \exp\left(-i\frac{z}{2}n\delta\right)\right|^2.
\end{equation}

Since the phases corresponding to different components are different, 
the components will interfere with each other and form Airy breathers \cite{driben.ol.19.5523.2014}.
However, if $zn\delta/2=2m\pi$ with $m$ being a nonzero integer, the phase contributions from all the components will be the same.
In other words, all the components will be in-phase,
because $\exp(iz/2\cdot n\delta)\equiv1$.
So, the intensity is the same as the input, except for the transverse displacement, due to the accelerating coordinate $x-z^2/4$.
This phenomenon is named the accelerating self-imaging or Airy-Talbot effect \cite{lumer.prl.115.013901.2015},
and the distance at which the first recurrence appears is called the Talbot length,
\begin{equation}\label{eq10}
  z_T=\frac{4\pi}{\delta}.
\end{equation}

\begin{figure}[htbp]
  \centering
  \includegraphics[width=0.5\columnwidth]{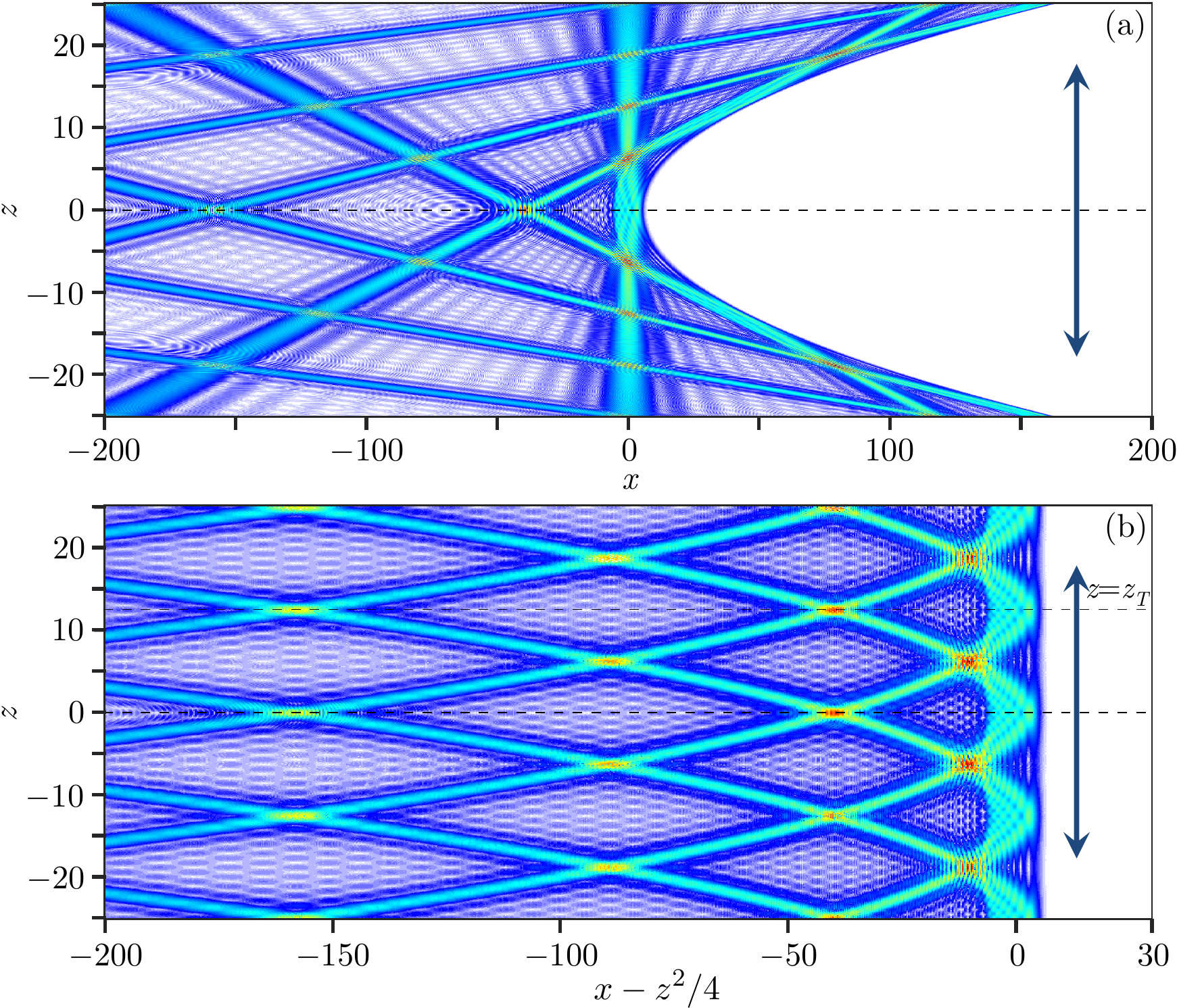}
  \caption{Airy-Talbot effect in (a) $(x,z)$ coordinates; (b) accelerating $(x-z^2/4,z)$ coordinates.
  Parameters are $\delta=1$, $c_n=1$ and $n\in[-5,~ 5]$.}
  \label{fig1}
\end{figure}

In Fig. \ref{fig1}, we display the intensity carpet of the Airy-Talbot effect in the real frame of reference ($z,x$) [Fig. \ref{fig1}(a)] and in the accelerating frame [Fig. \ref{fig1}(b)],
by setting $\delta=1$ and by choosing the components $c_n=[\cdots,1,1,1,\cdots]$, for $n\in[-5,~5]$.
To depict the accelerating property more clearly, we combine the positive
and negative propagation cases, separated by the dashed line and indicated by the arrows.
From Fig. \ref{fig1} it is evident that the self-imaging is more easily identified in the accelerating coordinates than in the real coordinates.
One easily recognizes the accelerating self-images at $\pm z_T$, $\pm 2z_T$, etc. 

In addition to $z_T$, another place of interest is at the half of the Talbot length,
where another constructive interference appears.
By definition, this distance is given by
$z_H=z_T/2$, but the image seen is not of the input beam, as can be seen in Fig. \ref{fig1}; in fact, from Eq. (\ref{eq9}), it is given by
\begin{equation}\label{eq9h}
	I_H(x,z)=\left|\sum_{n\in\mathbb{Z}}c_n {\rm Ai}\left(x-n\delta-\frac{z^2}{4}\right) \exp\left(-in\pi\right)\right|^2,
\end{equation}
which is similar to the input superposition, but with the redefined (sign-changing) coefficients $c'_n=c_n\exp(-in\pi)$.
One finds that $z_Hn\delta/2=(2m-1)\pi$ if $n$ is odd and $z_Hn\delta/2=2m\pi$ if $n$ is even.
That is, there is a $\pi$ phase shift between the $n^{\rm th}$ and $(n+1)^{\rm st}$ components and their images are identical.
Thus, the images at $z_H $ are as sharp as the ones at $z_T$, are phase-shifted relative to the input,
come from the sign-alternating input components,
and recur at the same intervals $\pm z_T$.
For this reason, we call the phenomenon {\it the dual Airy-Talbot effect}.

\begin{figure}[htbp]
	\centering
	\includegraphics[width=0.5\columnwidth]{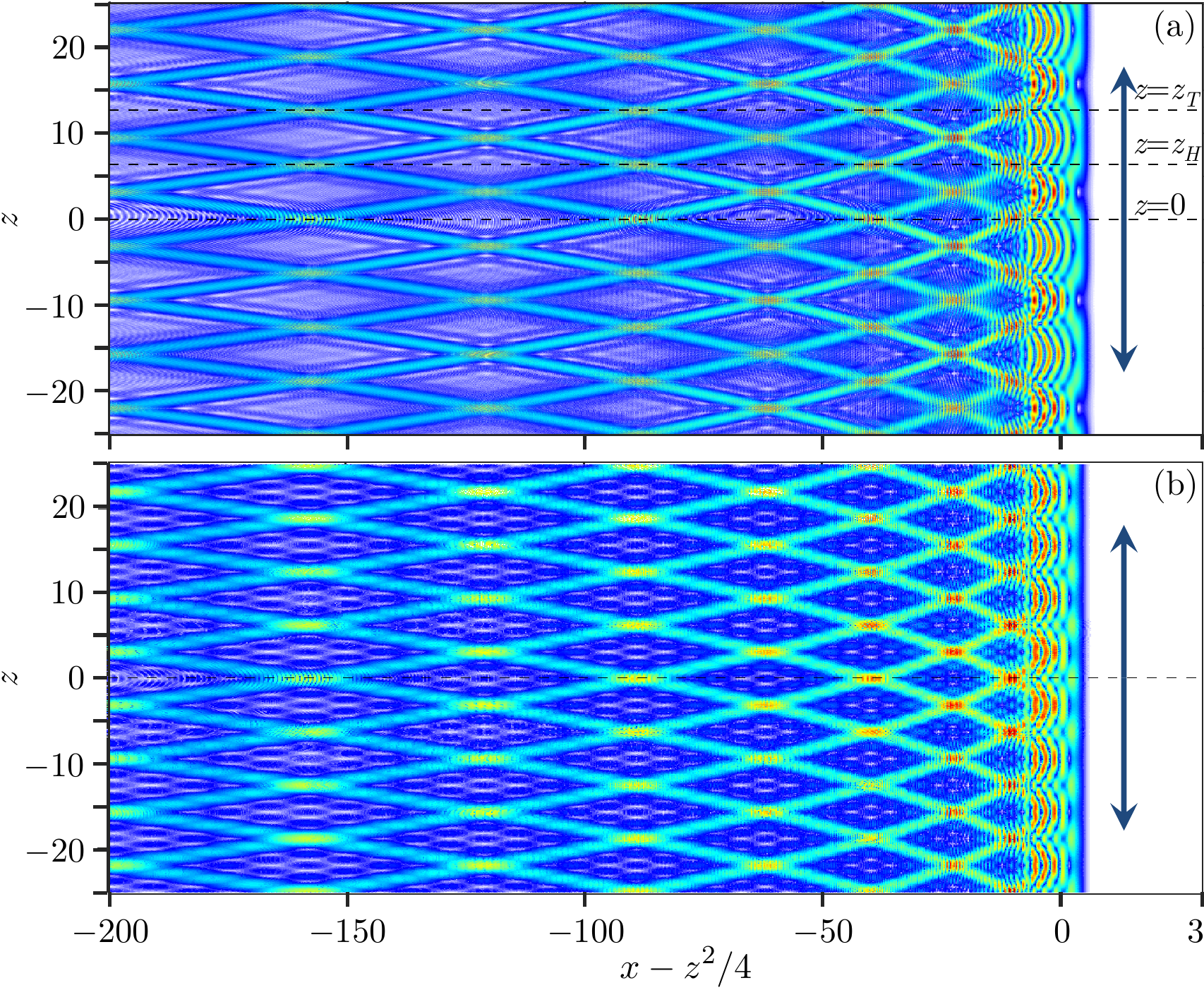}
	\caption{DualAiry-Talbot carpet in the accelerating frame.
            (a) The case with $c_n=[\cdots,1,i,1,i,1,\cdots]$.
            (b) The case with $c_n=[\cdots,1,0,1,0,1,\cdots]$.}
	\label{fig2}
\end{figure}

Different Talbot and dual-Talbot images can be constructed for different choices of the coefficients $c_n$.
For example, if we choose $c_n=[\cdots,1,i,1,i,1,\cdots]$, i.e.,
assume the coefficients of the odd components are $i$,
then Eq. (\ref{eq8}) can be recast into
\begin{align}\label{eq8b}
  \psi(x,z)=&\left[ \sum_{even} {\rm Ai}\left(x-n\delta-\frac{z^2}{4}\right) + i\sum_{odd} {\rm Ai}\left(x-n\delta-\frac{z^2}{4}\right) \right]
 \exp\left[i\frac{z}{2}\left(x-\frac{z^2}{6}\right)\right]
\end{align}
and
\begin{align}\label{eq8c}
  \psi(x,z)=\left[ \sum_{even} {\rm Ai}\left(x-n\delta-\frac{z^2}{4}\right) - i\sum_{odd} {\rm Ai}\left(x-n\delta-\frac{z^2}{4}\right) \right] 
   \exp\left[i\frac{z}{2}\left(x-\frac{z^2}{6}\right)\right]
\end{align}
at $z=z_T$ and $z=z_H$, respectively.
In Eqs. (\ref{eq8b}) and (\ref{eq8c}), the phase shifts between the two nearest components are $\pi/2$ and $-\pi/2$, respectively.
Thus, the amplitude (except for the exponential term) at $z_H$ is the  conjugate of that at $z_T$, but the intensities are the same.
Therefore, in this case the beam at $z_H$ -- the dual Airy-Talbot image -- is identical to the Airy-Talbot image at $z_T$.
We show this effect in Fig. \ref{fig2}(a).
The same result is obtained if $c_n=[\cdots,1,-i,1,-i,1,\cdots]$ is chosen.

To clarify things even further, we consider the case with $c_n=[\cdots,1,0,1,0,1,\cdots]$,
which makes the components in phase at $z_H$. The result is shown in Fig. \ref{fig2}(b).
One should note that the Talbot length here is halved, in comparison to that shown in Fig. \ref{fig1},
which can be easily seen if one replaces $n$ by $2n$ in Eq. (\ref{eq9}) --
making the problem identical to the one in Fig. \ref{fig1}.
As a result, $z_H$ mentioned above is the Talbot length for this case.
But, one still observes the dual Airy-Talbot effect at the half-Talbot length $z_H/2$, that is, at the quarter of the original Talbot length.

We would like to point that if there is only one component (e.g. $n=0$) in Eq. (\ref{eq8}),
the solution is reduced to the well-known Airy solution \cite{berry.ajp.47.264.1979}
\begin{equation}\label{eq12}
  \psi(x,z)={\rm Ai}\left(x-\frac{z^2}{4}\right) \exp\left[i\frac{z}{2}\left(x-\frac{z^2}{6}\right)\right].
\end{equation}
Thus, an Airy wave function itself can be considered just as an accelerating self-imaging beam,
which can be understood in two different ways:\\
(1) If the transverse displacement $\delta\rightarrow\infty$, the components will not affect each other.
Therefore, one can obtain a single Airy function as given in Eq. (\ref{eq12}), and the corresponding Talbot length is $z_T\rightarrow0$.\\
(2) If the transverse displacement $\delta\rightarrow0$, all the components will overlap into the Airy function given in Eq. (\ref{eq12}), and thus a single Airy function is obtained in the limit $z_T\rightarrow\infty$.

Based on the above analysis, the Airy function possesses the characteristics of duality
and indicates a close unity of the two opposite limits.
Therefore, the Airy function,  the Airy-Talbot effect, and the Airy breather can be considered from a unified point of view.
The appearance of the latter two phenomena results from the interference among Airy wave functions with different transverse displacements.
While the Airy-Talbot effect is the result of interference of many displaced Airy functions, the Airy breather is a result of the interference of just two \cite{driben.ol.19.5523.2014}.

\begin{figure*}
	\centering
	\includegraphics[width=\textwidth]{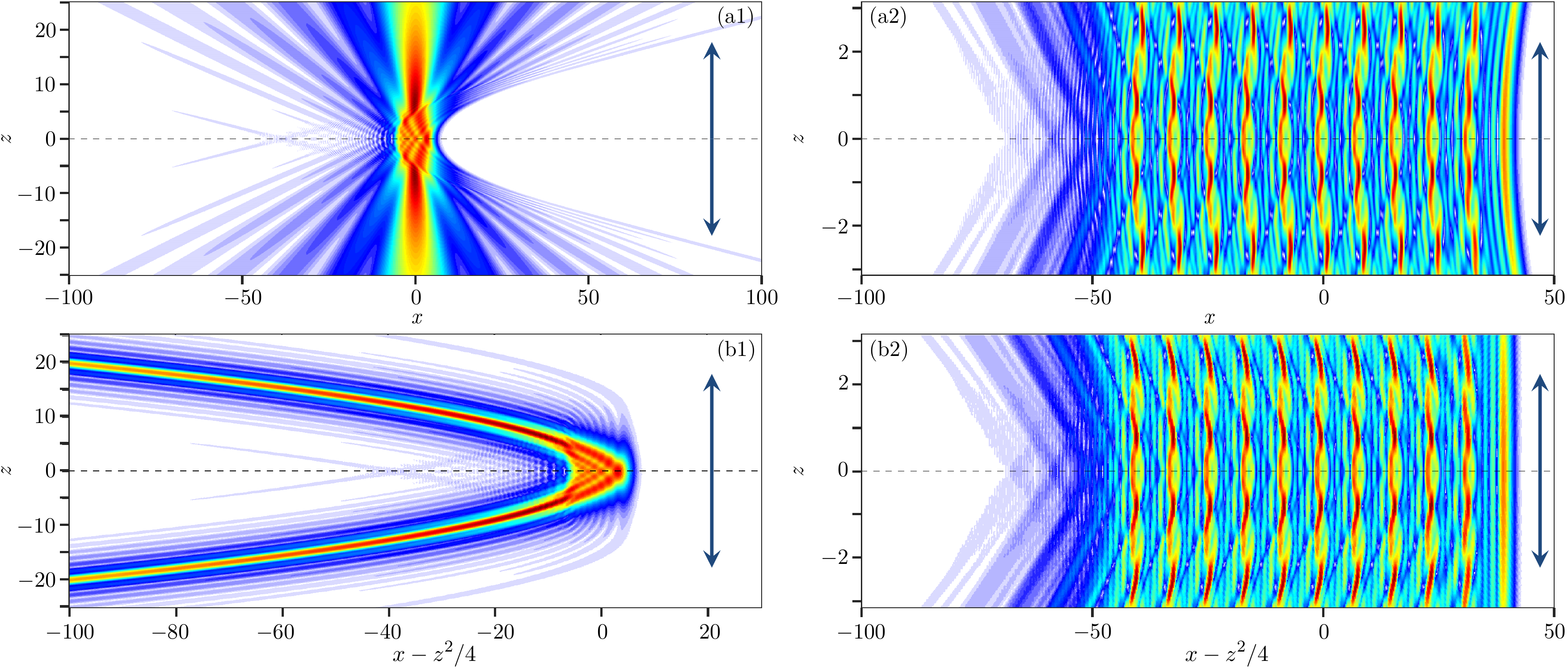}
	\caption{Same as Figs. \ref{fig1} and \ref{fig2}, but for the finite energy input with $a=0.1$. Here
		$\delta=1$ in the upper panels and $\delta=8$ in the lower panels.}
	\label{fig3}
\end{figure*}

The energy of all the above-considered Airy functions is infinite, which is not very realistic; however, in a similar vein a plane wave is also not realistic -- but still very useful.
As done earlier \cite{siviloglou.ol.32.979.2007,zhang.ol.38.4585.2013,zhang.oe.22.7160.2014,zhang.oe.23.10467.2015}, an apodization is needed.
Corresponding to Eq. (\ref{eq3}), the finite energy version can be written as
\begin{equation}\label{eq13}
  \psi(x)=\sum_{n\in\mathbb{Z}}c_n {\rm Ai}(x-n\delta)\exp[a(x-n\delta)],
\end{equation}
where $a$ is the decay factor.
Following the same procedure as above, the propagating solution with Eq. (\ref{eq13}) being the input is given by
%
\begin{align}\label{eq14}
  &\psi(x,z)=\sum_{n\in\mathbb{Z}}c_n {\rm Ai}\left(x-n\delta-\frac{z^2}{4}+iaz\right) \times \notag \\
  &\exp\left[i\frac{z}{2}\left(x-n\delta-\frac{z^2}{6}+a^2\right)\right] \exp\left[\left(x-n\delta-\frac{z^2}{2}\right)a\right],
\end{align}
which reduces to Eq. (\ref{eq8}) if $a=0$ and Eq. (5) in Ref. \cite{siviloglou.ol.32.979.2007} if only $n=0$ is considered.
The corresponding intensity is
\begin{align}\label{eq15}
  &I(x,z)=\exp\left[\left(2x-z^2\right)a\right] \times \notag \\
  & \left| \sum_{n\in\mathbb{Z}}c_n {\rm Ai}\left(x-n\delta-\frac{z^2}{4}+iaz\right) \exp\left[-\left(a+i\frac{z}{2}\right) n\delta \right] \right|^2,
\end{align}
in which the decay factor $a$ affects the Airy function profile during propagation and leads to the accelerating range being limited to only few Rayleigh lengths, so
that the self-imaging property will be affected even though $z n\delta/2=2m\pi$.
In Figs. \ref{fig3}(a1) and \ref{fig3}(b1), we display the propagation of a finite-energy input with $c_n=[\cdots,1,1,1,\cdots]$,
in which the Airy-Talbot effect disappears fast.
The reason is that the Talbot length is too long and beyond the accelerating range.
So, one has to enlarge $\delta$ (viz. shorten $z_T$), to make sure that the accelerating range is long enough for the formation of the Airy-Talbot effect.
In Figs. \ref{fig3}(a2) and \ref{fig3}(b2) we show the result with $\delta=8$, in which the Airy-Talbot effect can be seen,
even though there are some distortions, due to finite energy.

Another interesting phenomenon is that the beam propagates almost along straight lines when the propagation distance is long enough, as shown in Figs. \ref{fig3}(a1) and \ref{fig3}(b1).
As mentioned above, the accelerating range is only limited to a few Rayleigh lengths,
and with the increasing distance, each finite-energy Airy component is getting closer to a Gaussian-like beam \cite{siviloglou.ol.32.979.2007} with the symmetry axis $x=n\delta$.
As a result, the propagation is more and more similar to the interference of Gaussian-like beams, which do not possess accelerating property.

In summary, we have theoretically and numerically investigated the dual Airy-Talbot effect by superposing Airy wave functions with successively changing transverse displacement
and properly choosing the coefficients of the components.
At half Talbot length, the intensity of the beam may not be the same as the input beam, since it comes from a redefined initial superposition --
which is the dual Airy-Talbot effect.
We have also investigated the propagation of superposed finite-energy Airy functions.
Affected by the decay factor, the Airy-Talbot effect can only be obtained with large transverse displacements.
Last but not least, in our approach the concepts of an Airy breather and Airy-Talbot effect can be unified, because both stem from the interference of Airy wave functions.
The dual Airy-Talbot effect can have potential applications in particle capture,
light-bullet generation, photolithography, optical testing, and elsewhere.
In addition, it can be also used in florescence microscopy \cite{jia.np.8.302.2014}
if $c_n$ are chosen as multiplications of random numbers and $[\cdots,1,i,1,i,1,\cdots]$.

This work is supported by National Basic Research Program of China (2012CB921804),
National Natural Science Foundation of China (61308015, 11474228),
Key Scientific and Technological Innovation Team of Shaanxi Province (2014KCT-10),
and Qatar National Research Fund  (NPRP 6-021-1-005).

%

\end{document}